# Lateral beam shifts and depolarization upon oblique reflection from dielectric mirrors

Yuzhe Xiao[1], Linipun Phuttitarn[2], Trent Michael Graham[2], Chenghao Wan[1], Mark Saffman[2], and Mikhail A. Kats[1,2*]

[1]Department of Electrical and Computer Engineering, University of Wisconsin-Madison, Madison, Wisconsin 53706, USA
[2]Department of Physics, University of Wisconsin-Madison, Madison, Wisconsin 53706, USA
*Correspondence to: mkats@wisc.edu

**Abstract:** Dielectric mirrors comprising thin-film multilayers are widely used in optical experiments because they can achieve substantially higher reflectance compared to metal mirrors. Here we investigate potential problems that can arise when dielectric mirrors are used at oblique incidence, in particular for focused beams. We found that light beams reflected from dielectric mirrors can experience lateral beam shifts, beam-shape distortion, and depolarization, and these effects have a strong dependence on wavelength, incident angle, and incident polarization. Because vendors of dielectric mirrors typically do not share the particular layer structure of their products, we designed and simulated several dielectric-mirror stacks, and then also measured the lateral beam shift from two commercial dielectric mirrors and one coated metal mirror. We hope that this paper brings awareness of the tradeoffs between dielectric mirrors and front-surface metal mirrors in certain optics experiments, and suggest that vendors of dielectric mirrors provide information about beam shifts, distortion, and depolarization when their products are used at oblique incidence.

Dielectric mirrors, comprising multilayers of different dielectric materials, can achieve very high reflectance spanning some spectral range, often much higher than the reflectance of conventional metal mirrors [1]. However, because the reflectance of metal mirrors is usually a result of a single-surface reflection, whereas dielectric mirrors function on the principle of thin-film interference, the reflectance of dielectric mirrors can have a much larger dependence on the angle of incidence. In addition to potential reductions in reflectance at some angles, beams reflected from a dielectric mirror may also undergo lateral shifts, distortion, and depolarization. In particular, the lateral shift of beams reflected from various photonic structures has been an active research topic, and is regularly seen in reflection from metallic surface [2], dielectric slabs [3], photonic crystals [4], metal-dielectric multilayers [5], among others. At a single interface, the lateral beam shift is known as the Goos-Hänchen shift [6], [7], though the term is also sometimes extended to describe lateral shifts in the aforementioned photonic structures [3]–[5], [8].

Despite the broad use of dielectric mirrors in optical setups, their effects on the profile of the reflected beam have not been widely appreciated. Though they may tailor mirror designs to a particular application, most commercial vendors of dielectric mirrors do not directly provide information about the angle- and wavelength-dependent lateral shift, beam distortion, and depolarization of their dielectric mirrors, and the precise layered designs are typically trade secrets, so customers cannot perform their own calculations. Indeed, the phenomena reported here were encountered unexpectedly during assembly of a complex system for optical control of atomic qubits [9] and necessitated adjusting the optical layout to minimize the polarization-dependent beam shifts.

In this article, we provide analysis of the lateral beam shift, beam-shape distortion, and depolarization when light is reflected from various dielectric mirrors at oblique incident angles. We look at two examples: a linearly chirped Bragg mirror, and a more-sophisticated dielectric mirror designed using thin-film optimization. We also experimentally demonstrate these effects in two commercial dielectric mirrors. We hope that this article can bring awareness to potential issues when using dielectric mirrors at oblique incident angles and suggest that vendors of dielectric mirrors provide detailed information about the performance of their product at oblique incidence.

The most straightforward design of a dielectric mirror is a Bragg mirror which consists of alternating quarter-wave layers of high- and low-index dielectrics. Due to interference, Bragg mirror has a spectral range of high reflectance ("photonic band gap") determined by the index contrast [1], [10]. To increase the spectral range, and thus make a broadband dielectric mirror, one can chirp the Bragg grating such that the period varies gradually as a function of depth [11]. This way, light at different wavelengths is reflected at different positions in the thin-film stack. One example of a chirped Bragg grating is illustrated in Fig. 1(a), comprising 20 alternating layers of low- and high-index ($n_1 = 1.5, n_2 = 2.5$) dielectrics with a total thickness of about 2 µm. Thicknesses for each layer are listed in Table 1 in the Appendix. For simplicity, we assume no dispersion. In this mirror, the Bragg wavelength gradually changes from 0.6 µm to 1 µm from the top to the bottom (schematics in Fig. 1(a)), resulting in high reflectance for wavelengths from 0.7 µm to about 1 µm at normal incidence (Fig. 1(b)).

When light is incident at an oblique angle on this chirped Bragg mirror, the short-wavelength component will be reflected from the top few layers with minimal lateral beam shift, while the long-wavelength component can reach deeper into the mirror and then emerges from the top surface of the mirror with a considerable lateral shift (e.g., at $\lambda_3$ in Fig. 1(a)). In addition to the dependence on wavelength, the lateral beam shift also depends on the polarization state of the incident light. As illustrated in Fig. 1(a), p-polarized light generally penetrates deeper into the mirror due to the Brewster effect [12] and hence typically experiences a larger lateral beam shift than s-polarized light. Note that this situation can sometimes be reversed with larger shifts for s-polarized light, as discussed later in this article.

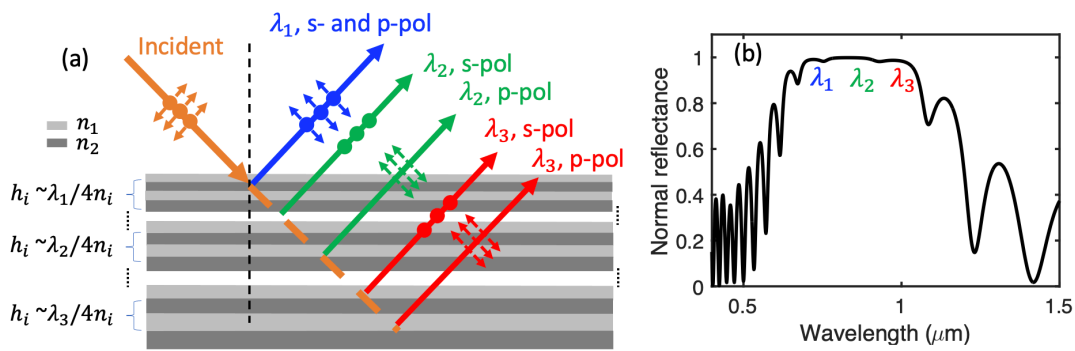

**Figure 1**. **Lateral beam shift from a chirped Bragg mirror.** (a) Schematic of a chirped Bragg mirror designed for high reflectance from 0.7 µm to 1 µm. The mirror consists of 20 alternating quarter-wave layers of low ($n_1 = 1.5$) and high ($n_2 = 2.5$) index dielectrics, with a total thickness about 2 µm. The mirror is chirped such that the Bragg wavelength changes linearly from 0.6 µm to 1 µm from the top to the bottom. Light with different wavelengths and for different polarizations is reflected at different depths in the mirror, leading to wavelength- and polarization-dependent lateral beam shift at oblique incidence. (b) Normal-incidence reflection spectrum of the Bragg mirror in (a).

For a more quantitative picture of the lateral beam shift, we calculated the reflection of a Gaussian beam at oblique incidence from the dielectric mirror in Fig. 1. As an illustration, we first consider the easier-to-calculate two-dimensional (2D) case, assuming the Gaussian beam propagates in the $x - z$ plane and is invariant along the $y$ axis. Later in the manuscript, we also analyze the 3D case, and the results are similar to the 2D case. The calculation is based on the plane-wave decomposition ("angular spectrum") method [13], [14], where for each plane wave we calculate the reflectance using the transfer-matrix method [15].

Because we can decompose any incident light beam into s- and p-polarized components, the process described below is performed individually for s- or p-polarized monochromatic light.

The numerical approach for calculating the beam shift and other effects given an arbitrary incident field $E_{in}(x)$ defined just above the surface of the mirror is schematically shown in Fig. 2(a). We first calculate the angular spectrum of the incident beam, $A_{in}(k_x)$, where $k_x$ is the parallel component of the wavevector, by taking a Fourier transform of $E_{in}(x)$ over the spatial coordinate $x$:

$$A_{in}(k_x) = \int_{-\infty}^{\infty} E_{in}(x) e^{ik_x x} dx \quad (1)$$

The reflected angular spectrum, $A_{out}(k_x)$, is related to $A_{in}(k_x)$ via the wavelength- and polarization-dependent Fresnel reflection coefficients of the dielectric mirror at various incident angles, $r(k_x)$:

$$A_{out}(k_x) = A_{in}(k_x) r(k_x) \quad (2)$$

The angle of incidence $\theta$ is related to the parallel wavevector as: $k_x = k_0 \sin(\theta)$, where $k_0$ is the wavenumber in free space. Finally, the reflected beam profile just above the mirror surface, $E_{out}(x)$, is obtained by taking the inverse Fourier transform of $A_{out}(k_x)$:

$$E_{out}(x) = \int_{-\infty}^{\infty} A_{out}(k_x) e^{-ik_x x} dk_x \quad (3)$$

This approach is general and can be applied to any arbitrary incident field. Let's first consider the simplest case of a plane wave, which has infinite spatial extent. The electric field for a plane wave propagating along the $x - z$ plane at an angle $\theta$ with the $x$ axis is

$$E_{in}(x, z) = e^{ik_0 \sin(\theta) x + ik_0 \cos(\theta) z} \quad (4)$$

The incident angular spectrum is simply a Dirac delta function $A_{in}(k_x) = \delta(k_x - k_0 \sin(\theta))$, only containing the wave vector along the propagation direction such that $k_x = k_0 \sin(\theta)$. It follows Eqs. 2 and 3 that the reflected beam will be the incident plane wave multiplied by a single Fresnel reflection coefficient of the mirror $r(k_0 \sin(\theta))$:

$$E_{out}(x, z) = r(k_0 \sin(\theta)) e^{ik_0 \sin(\theta) x - ik_0 \cos(\theta) z} \quad (5)$$

Because of the infinite spatial extent of the plane wave, there can be no beam shift.

However, any real finite-sized beam includes wave vectors along different directions (i.e., different values of $k_x$). These wave vectors are reflected by the mirror with different reflection coefficients $r(k_x)$, which can result in lateral shifts and beam-shape distortion depending on the response of the dielectric mirror (i.e., how $r$ changes vs. the incident angle). These effects become more pronounced for smaller (e.g., focused) beams. In the following examples, we use a focused near-infrared Gaussian beam with a beam size of a few µm to illustrate these effects.

Using Eqs. (1-3), we performed calculations for a Gaussian beam with full width at half maximum beam width of 3.5 µm and incident angle of 30° at two different wavelengths of 700 and 1000 nm, corresponding to the left and right edge of the spectral window of the dielectric mirror shown in Fig. 1. The calculation results are summarized in Figs. 2(b-c). We also performed full-wave simulations using the finite-difference time-domain (FDTD) method (implemented using Ansys Lumerical FDTD) to verify our calculation and to directly visualize the field profile inside the dielectric mirror. The intensity distributions are summarized in Figs. 2(d-e). In the FDTD simulations, the Gaussian beam is injected at the top surface of the mirror at $x = z = 0$ µm and initially propagates along the $+z$ and $+x$ direction.

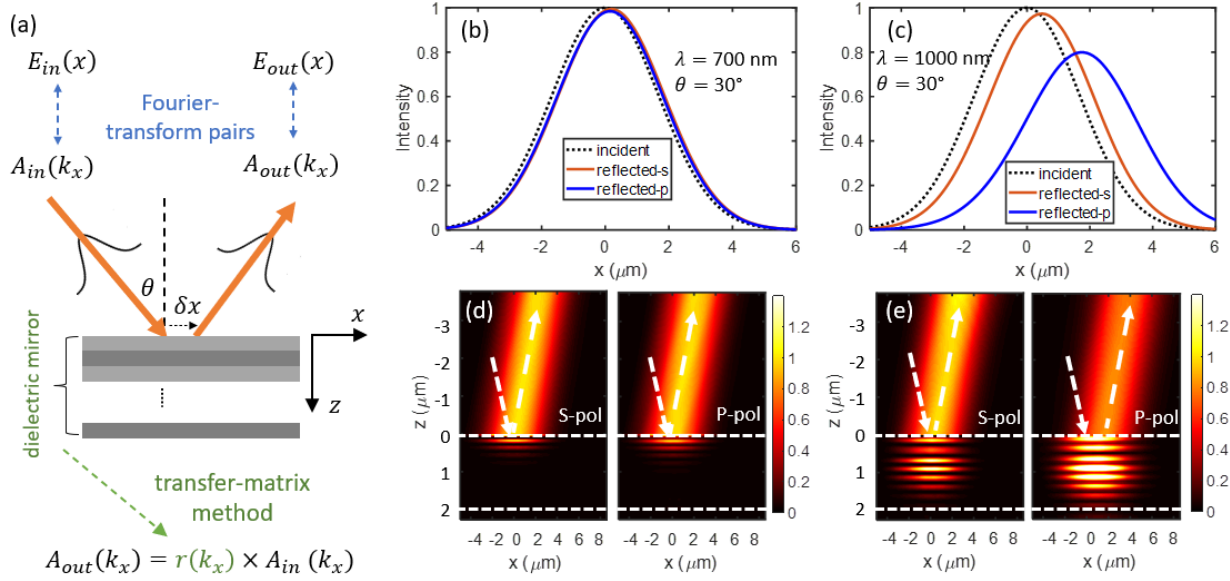

**Figure 2**. Simulation of the lateral beam shift from a chirped Bragg mirror. **(a)** Schematic of the numerical approach. **(b-c)** Calculated reflected beam profile for a Gaussian beam at an incident angle of 30° from the chirped Bragg mirror shown in Fig. 1 at wavelengths of (b) 700 and (c) 1000 nm. **(d-e)** Intensity distributions obtained via full-wave simulation of a Gaussian beam obliquely incident on the chirped Bragg mirror at wavelength of (c) 700 and (d) 1000 nm. The chirped Bragg mirror region is outlined by the white-dashed lines.

At $\lambda = 700$ nm, light is efficiently reflected by the very top portion of the dielectric mirror. Therefore, reflected beams for both p- and s-polarizations almost overlap with the incident beam, showing minimum lateral beam shift (Fig. 2(b)), and the reflectance is close to 1. The field profiles obtained via full-wave simulation show that only the top portion of the dielectric mirror is excited with the reflected beam centered around $x = 0$ μm for both polarizations (Fig. 2(d)).

For $\lambda = 1000$ nm, the p-polarized light beam experiences a lateral shift about 2 μm, while the shift is much smaller (<0.5 μm) for s-polarization (Fig. 2(c)). In addition, the peak intensity of the p-polarized light beam is reduced by about 20% with a slightly broadened beam size. Later, we calculated in Fig. 3(d) the reflectance for this case to be 86%. Therefore, most of the drop in peak intensity is due to transmission while only a small fraction (< 25%) is due to broadening. The full-wave simulation shows that almost the whole dielectric stack is excited by light at $\lambda = 1000$ nm, resulting in a much larger lateral shift (about 2 μm) for the p-polarized beam. In this case, the reflectance is slightly lower for p-polarization (also as seen in Fig. 2(c), though the reflectance can be increased back up to ~1 by increasing the thickness of the mirror and changing the "period" more gradually.

We note that a considerable lateral shift only occurs at certain combinations of wavelength, incident angle and polarization state. To quantify such effects, we calculate the lateral beam shift $\delta x$ (with respect to an incident beam centered at $x = 0$ μm) and reflectance $R$ as follows

$$\delta x = \frac{\int_{-\infty}^{\infty} x |E_{out}(x)|^2 dx}{\int_{-\infty}^{\infty} |E_{out}(x)|^2 dx} \tag{6}$$

$$R = \frac{\int_{-\infty}^{\infty} |E_{out}(x)|^2 dx}{\int_{-\infty}^{\infty} |E_{in}(x)|^2 dx} \tag{7}$$

Then we calculate the lateral shift and reflectance for all possible combinations of incident angle, wavelength, and polarization, and summarize the results in Fig. 3. Substantial beam shifts can be observed for certain combinations of incident angle, wavelength, and polarization. Note that the reflectance reduction for the p-polarized light at long wavelengths (> 950 nm) and large angles ($\theta > 40°$) is due to the finite thickness of mirror.

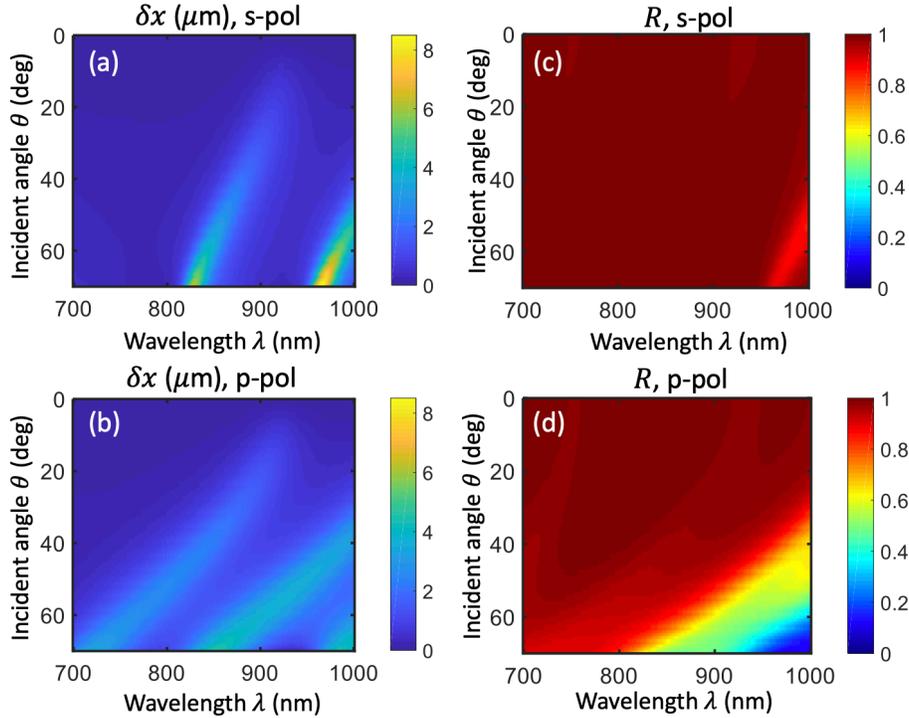

**Figure 3. (a and b)** Polarization-dependent lateral beam shift and **(c and d)** reflectance as a function of incident wavelength and angle, for a Gaussian beam and the dielectric mirror shown in Fig. 1. The incident Gaussian beam has a beam width (full width at half maximum) of about 3.5 μm.

The example shown in Figs. (1-3) is based on a linearly chirped Bragg mirror and assumes no dispersion for the dielectric materials. To extend the discussion to realistic dielectric mirrors, we designed a mirror stack using the open design tool from LightMachinery Inc [16], consisting of 25 layers of $TiO_2$ and $SiO_2$, resulting in high normal-incidence reflectance from 450 to 800 nm (Fig. 4(a)). The thicknesses for each layer are listed in Table 2 in the Appendix.

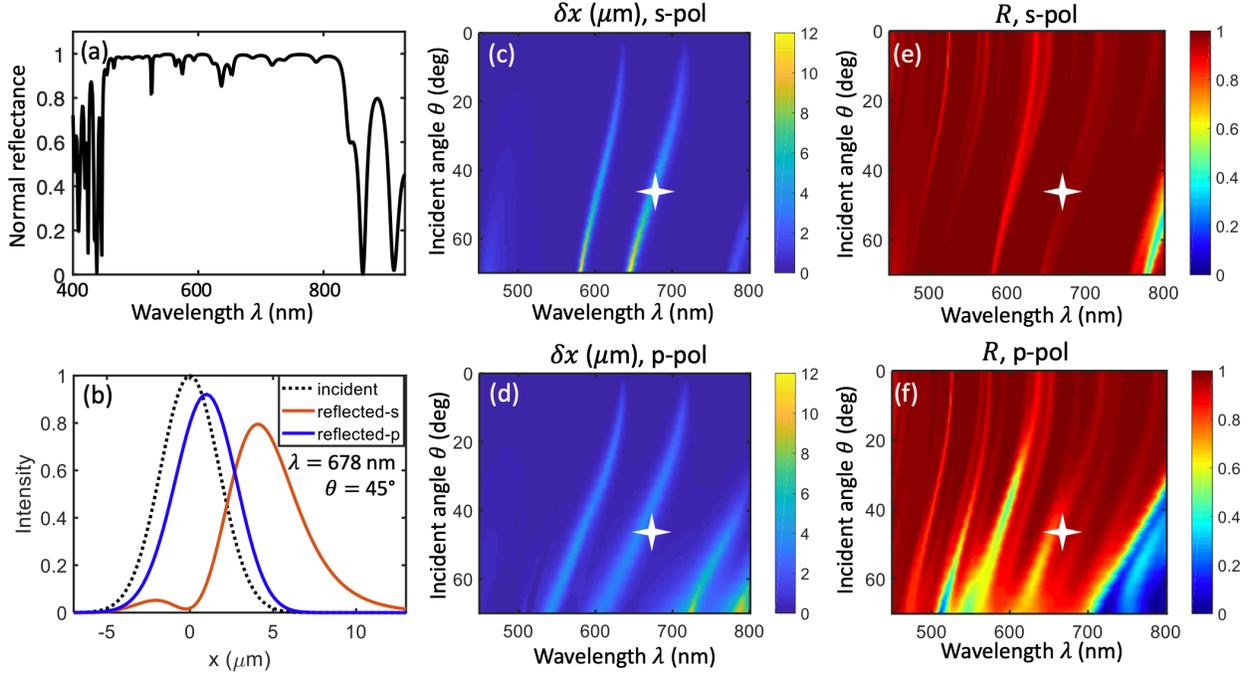

**Figure 4. Simulated lateral shift and beam-shape distortion from a realistic dielectric mirror. (a).** Normal-incidence reflectance of a dielectric mirror designed for 450 and 800 nm, based on 25 alternating layers of $TiO_2$ (high-index coating) and $SiO_2$ (low-index coating). Thicknesses for each layer are listed in Table 2 in Appendix. **(b)** Position-dependent intensity distributions for the incident and reflected beams for two polarizations, for an incident Gaussian beam ($\theta = 45°$) at $\lambda = 678$ nm with a beam width of about 3.5 µm. The polarization-dependent lateral shift and distortion is visible. **(c-f)** Maps of (c-d) the beam shift $\delta x$ and (e-f) reflectance $R$ as a function of the incident angle and wavelength.

Compared to the response of a linearly chirped Bragg mirror (Fig. 3), this dielectric mirror has many more modes (Figs. 4(e and f)), resulting in considerable beam shifts (Figs. 4(c and d)). As one example, in Fig. 4(b) we show the calculated intensity profiles for reflected s- and p-polarized beams for $\lambda = 678$ nm and $\theta = 45°$. Both polarizations show considerable lateral beam shift; especially, a lateral beam shift larger than the beamwidth is observed for the s-polarization. In addition to the lateral beam shift, s-polarized beam also exhibits considerable beam-shape distortion: it is no longer a simple Gaussian beam (Fig. 4(b), orange curve).

Note that, in general, we expect p-polarized light to penetrate deeper into the thin-film stack due to the Brewster effect, therefore resulting in a larger lateral shift. However, as seen in Fig. 4, the resonance modes of a complex thin-film assembly tend to be sharper for s-polarized light due to stronger interface reflections, and in the vicinity of these modes the lateral shift for s-polarized light can be significantly larger than for p-polarized light. This phenomenon can also be seen in Figs. 3(a and b).

The situation can become more complicated for a light beam with a polarization state other than s or p. As an illustration, we show in Fig. 5 the simulation results for the reflection of a right-circularly-polarized (RCP) focused Gaussian beam from the dielectric mirror discussed in Fig. 4. The incident beam is at $\lambda = 678$ nm and incident angle $\theta = 45°$, which is the same scenario as in Fig. 4(b), though now we switch to full 3D calculations. As shown in Fig. 5(a) and (b), the reflected beam is stretched due to the different lateral shift between p- and s-polarized components of the electric field.

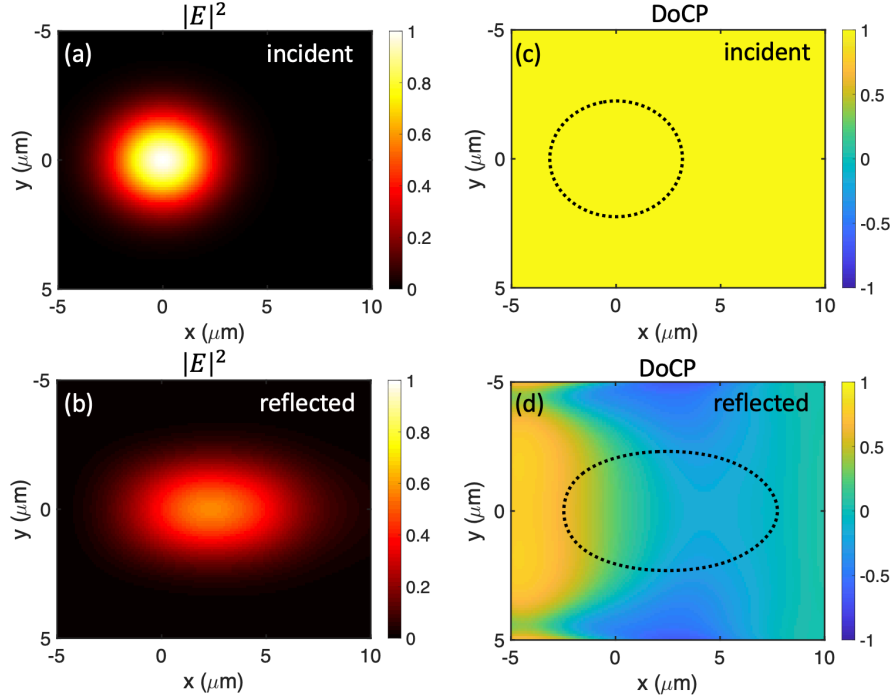

**Figure 5**. **Simulated beam-shape distortion and polarization change of a focused circularly polarized Gaussian beam from the dielectric mirror shown in Fig. 4.** The Gaussian beam at 678 nm is right circularly polarized (RCP) and incident at an angle of 45° (same as in Fig. 4(d)). (**a** and **b**): Intensity profile of the incident and reflected Gaussian beam. Degree of circular polarization (DoCP) for the incident (**c**) and reflected beam (**d**). The dotted circles in (c) and (d) represent the contour line of 20% of the peak intensity of the beam. Note that the intensity and DoCP of the reflected beam is presented immediately above the top surface of the mirror, with no additional propagation.

To quantify the depolarization upon reflection, we plot the degree of circular polarization (DoCP) for the incident and reflected beams in Fig. 5 (c) and (d). DoCP is related to the Stokes parameters, $S_1$ and $S_3$, of the electric field as [17]:

$$\text{DoCP} = \frac{S_3}{S_1} = \frac{2E_x E_y \sin\Delta}{E_x^2 + E_y^2} \tag{8}$$

where $E_x$ and $E_y$ are the amplitude of the electric field polarized along $x$ (here corresponds to p-polarization) and $y$ (here corresponds to s-polarization), and $\Delta$ is the phase difference between $E_x$ and $E_y$. The incident beam is RCP, with DoCP = 1 everywhere. Due to the polarization-dependent response of the dielectric mirror, the polarization of the reflected beam is modified significantly and has position-dependent DoCP across the beam.

We measured the lateral beam shift from two commercial broadband dielectric mirrors (Thorlabs BB05-e03 and New Focus 5102) and one silver mirror with a protective dielectric coating (Newport 5153). The schematic of the experimental setup is shown in Fig. 6(a). Incident light from a 1040 nm continuous-wave laser (Time-Base ECQDL-200F) is linearly polarized using a polarized beam splitter and then focused down to a beam size of about 18 μm onto the mirrors with an incident angle of 45°. We varied the polarization angle of the incident light by rotating a half-wave plate. The mirror is mounted on a rear-loaded mirror mount where the mirror is inserted from the back and the mirror reflective surface is pushed against the mounting stop surface. This mounting mechanism allows the reflective surfaces of different mirrors to be fixed against the same stop surface regardless of the mirror thickness and enable the mirror to be switch without affecting the focus. The beam reflected from the mirror is then magnified 8 times onto the beam profiler (WinCamD-XHR) using 40 mm and 500 mm lenses, with the final beam size of about 220 μm. For each polarization setting, the result was measured 1000 times and averaged. The lateral beam shift is then determined by the shift relative to the position of the beam for p-polarization, which is shown in Figure 6(b) for three mirrors as a function of polarization angle. The Newport 5153 metal mirror produced the smallest lateral shift compared to the dielectric mirrors, though note that the protective coating on the mirror likely increases the lateral shift beyond that of an unprotected metallic mirror. Comparing the two dielectric mirrors, Thorlabs BB05-e03 has a lateral shift 3.2 times larger than that of the New Focus 5102 mirror. Since the layer designs are not provided by the manufacturers, we were not able to simulate the beam shifts for these three specific mirrors.

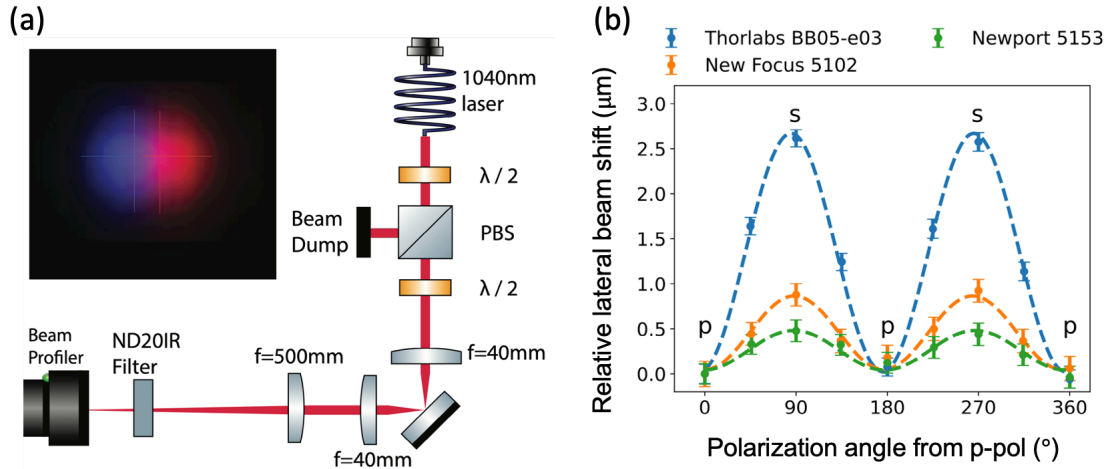

**Figure 6. Experimental measurements of lateral beam shift from commercial mirrors.** (a): Schematic of the experimental setup. Linearly polarized light is focused on the mirror, and the lateral beam shift is measured. The inset shows the spots on the beam profiler, for p-polarization (blue) and s-polarization (red). (b) Measured relative lateral beam shift as a function of the angle of polarization; "relative" means that we plot the beam shift with respect to the beam position for p-polarized input. We measured two commercial near-infrared broadband dielectric mirrors (Thorlabs BB05-e03 and New Focus 5102) and one coated metal mirror (Newport 5153 silver mirror). The incident light is a focused Gaussian beam $\lambda = 1040$ nm, a beam size of 18 μm, and an incident angle of 45°.

Note that the lateral beam shift seems to not depend much on the size of the incident beam (Fig. A1 in the Appendix). At the same time, for wider beams, the beam shift can effectively become negligible, thereby reducing the beam-shape distortion. At visible and near-infrared frequencies, these effects can usually be safely ignored for light beams wider than hundreds of microns. On the other hand, for tightly focused light beams at oblique incidence, we recommend either using an uncoated front-surface metal mirror, or confirming beforehand that the dielectric mirror will not lead to issues at the particular wavelength and incidence angle of a given experiment. We do note that depolarization may occur even for wider beams due to a polarization-dependent reflectance (both amplitude and phase).

To conclude, we found that the use of dielectric mirrors for focused light at oblique incidence can result in substantial lateral beam shifts (as large as ~10 μm for typical near-infrared mirrors), beam-shape distortion, and depolarization. These effects are not well-known, and cannot be easily simulated for commercial dielectric mirrors, because specific thin-film layer structures are typically trade secrets. We recommend that vendors make prospective end users aware of the risks of dielectric mirrors at oblique incidence, and provide additional details about their performance, such as maps of lateral shifts as a function the wavelength and angle of incidence for each of their products.

This material is based upon work supported by Q-NEXT, one of the U.S. Department of Energy Office of Science National Quantum Information Science Research Centers, and by NSF Award 2016136 for the QLCI center Hybrid Quantum Architectures and Networks and NSF Award 2210437

**Appendix:**

Table 1: thickness of the 20 quarter-wave layers in the dielectric mirror in Fig. 1(a). The odd-number layers have refractive index of 1.5 and even-number layers have refractive index of 2.5.

| Layer # | 1 | 2 | 3 | 4 | 5 | 6 | 7 | 8 | 9 | 10 |
|---|---|---|---|---|---|---|---|---|---|---|
| Thickness (nm) | 100.0 | 60.0 | 107.4 | 64.4 | 114.8 | 68.9 | 122.2 | 73.3 | 129.6 | 77.7 |
| Layer # | 11 | 12 | 13 | 14 | 15 | 16 | 17 | 18 | 19 | 20 |
| Thickness (nm) | 137.0 | 82.2 | 144.4 | 86.7 | 151.2 | 91.1 | 159.3 | 95.6 | 166.7 | 100.0 |

Table 2: thickness of the 25 layers in the dielectric mirror in Fig. 3(a). The odd-number layers are $TiO_2$ and even-number layers are $SiO_2$.

| Layer # | 1 | 2 | 3 | 4 | 5 | 6 | 7 | 8 | 9 | 10 | 11 | 12 | 13 |
|---|---|---|---|---|---|---|---|---|---|---|---|---|---|
| Thickness (nm) | 91.1 | 83.0 | 74.9 | 64.1 | 45.9 | 60.1 | 69.5 | 72.9 | 68.9 | 60.1 | 50.0 | 64.8 | 72.2 |
| Layer # | 14 | 15 | 16 | 17 | 18 | 19 | 20 | 21 | 22 | 23 | 24 | 25 | |
| Thickness (nm) | 81.0 | 105.3 | 87.1 | 79.0 | 84.4 | 99.9 | 87.1 | 79.7 | 85.1 | 254.5 | 830.2 | 1371.6 | |

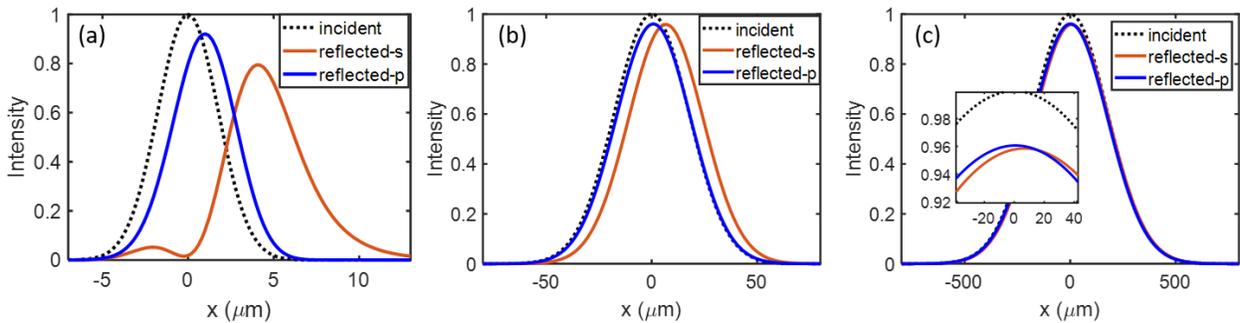

**Figure A1. Calculated reflected beam profiles for different-sized Gaussian beams.** All calculations are performed for wavelength of 679 nm and incident angle of 45° onto the dielectric mirror in Fig. 4. (a-c) Intensity of the incident and reflected beams just above the surface of the mirror as a function of the lateral position. (a) Replica of Fig. 4(b). (b, c) Calculation results for a Gaussian beam that is (b) 10 and (c) 100 times larger than the Gaussian beam in (a). The absolute values of the lateral beam shift are similar in all three cases.

**References:**

[1] H. A. Macleod, *Thin-Film Optical Filters*. CRC Press, 2021.
[2] M. Merano, A. Aiello, G. W. 't Hooft, M. P. van Exter, E. R. Eliel, and J. P. Woerdman, "Observation of Goos-Hänchen shifts in metallic reflection," *Opt. Express*, vol. 15, no. 24, pp. 15928–15934, Nov. 2007.
[3] H. Chen, L.-G. Wang, and S.-Y. Zhu, "Large negative Goos–Hänchen shift from a weakly absorbing dielectric slab," *Opt. Lett.*, vol. 30, no. 21, pp. 2936–2938, Nov. 2005.
[4] I. V. Soboleva, V. V. Moskalenko, and A. A. Fedyanin, "Giant Goos-Hänchen effect and Fano resonance at photonic crystal surfaces," *Phys. Rev. Lett.*, vol. 108, no. 12, p. 123901, Mar. 2012.
[5] H. Saito, M. Tomita, T. Matsumoto, and Y. Neo, "Giant and highly reflective Goos-Hanchen shift in a

metal-dielectric multilayer Fano structure," *Opt. Express*, vol. 27, no. 20, pp. 28629–28639, Sep. 2019.

[6]  F. Goos and H. Hänchen, "Ein neuer und fundamentaler Versuch zur Totalreflexion," *Ann. Phys.*, vol. 436, no. 7–8, pp. 333–346, Jan. 1947.

[7]  J. D. Love and A. W. Snyder, "Goos-Hänchen shift," *Appl. Opt.*, vol. 15, no. 1, pp. 236–238, Jan. 1976.

[8]  B. Wang, D. Zhao, P. Lu, Q. Liu, and S. Ke, "Giant Goos-Hanchen shifts in non-Hermitian dielectric multilayers incorporated with graphene," *Opt. Express*, vol. 26, no. 3, pp. 2817–2828, Feb. 2018.

[9]  T. M. Graham *et al.*, "Multi-qubit entanglement and algorithms on a neutral-atom quantum computer," *Nature*, vol. 604, no. 7906, pp. 457–462, Apr. 2022.

[10] P. Yeh, *Optical Waves in Layered Media*. Wiley, 2005.

[11] L. B. Glebov *et al.*, "Volume-chirped Bragg gratings: monolithic components for stretching and compression of ultrashort laser pulses," *Opt. Eng.*, vol. 53, no. 5, p. 051514, Feb. 2014.

[12] D. Brewster, "On the laws which regulate the polarisation of light by reflexion from transparent bodies," *Philos. Trans. R. Soc. London*, vol. 105, pp. 125–159, Dec. 1815.

[13] V. Shah and T. Tamir, "Absorption and lateral shift of beams incident upon lossy multilayered media," *JOSA*, vol. 73, no. 1, pp. 37–44, Jan. 1983.

[14] J. W. Goodman, *Introduction to Fourier Optics*. Roberts and Company Publishers, 2005.

[15] S. J. Byrnes, "Multilayer optical calculations," Mar. 2016.

[16] "Thin Film Cloud | LightMachinery." [Online]. Available: https://lightmachinery.com/optical-design-center/thin-film-cloud/.

[17] B. DeBoo, J. Sasian, and R. Chipman, "Degree of polarization surfaces and maps for analysis of depolarization," *Opt. Express*, vol. 12, no. 20, pp. 4941–4958, Oct. 2004.